\documentclass[twocolumn,preprintnumbers,superscriptaddress]{revtex4}
\usepackage{mathrsfs}
\usepackage{amssymb}
\usepackage{graphicx}
\usepackage{dcolumn}
\usepackage{bm}
\usepackage{graphicx}
\usepackage{float}
\usepackage{longtable}
\usepackage{amsmath}
\usepackage{multirow}
\usepackage{setspace}
\usepackage{color}

\newfont{\largemi}{cmmi10}
\newfont{\smallmi}{cmmi6}

\draft

\def\eqref#1{Eq.~(\ref{eq:#1})}

\begin{document}

\title{Unitary limit and linear scaling of neutrons in\\
 harmonic trap with tuned CD-Bonn and square-well interactions}

\author{Yi-Yuan Cheng \footnote{yycheng@phy.ecnu.edu.cn}}
\affiliation{Department of Physics, East China Normal University, Shanghai 200241, China}

\author{Thomas T. S. Kuo \footnote{kuo@tonic.physics.sunysb.edu}}
\affiliation{Department of Physics and Astronomy, Stony Brook University, New York 11794-3800, USA}

\author{Ruprecht Machleidt \footnote{machleid@uidaho.edu}}
\affiliation{Department of Physics, University of Idaho, Moscow, ID 83844, USA}

\author{Yu-Min Zhao \footnote{ymzhao@sjtu.edu.cn}}
\affiliation{School of Physics and Astronomy, Shanghai Jiao Tong University, Shanghai 200240, China}

\date{\today}

\begin{abstract}
We study systems of finite-number neutrons in a
harmonic trap at the unitary limit.
Two very different types of neutron-neutron interactions
are applied, namely, the meson-theoretic CD-Bonn potential
and hard-core square-well interactions,
all tuned to possess infinite scattering lengths,
and with effective ranges comparable to or larger than the trap size.
The potentials are renormalized
to equivalent, scattering-length preserving low-momentum
potentials, $V_{{\rm low}-k}$,
with which the particle-particle hole-hole
ring diagrams are summed to all orders
to yield the ground-state energy $E_0$ of the finite neutron system.
We find the ratio
$E_0/E_0^{\rm free}$ (where $E_0^{\rm free}$ denotes
the ground-state energy of the corresponding non-interacting system)
to be remarkably independent from variations of the
harmonic trap parameter, the number of neutrons,
the decimation momentum of $V_{{\rm low}-k}$,
and the type and effective range of the unitarity potential.
Our results support a special virial linear scaling relation of $E_0$.
Certain properties of Landau's quasi-particles
for trapped  neutrons at the unitary limit are also discussed.
\end{abstract}


\vspace{0.2in}

\maketitle

{\bf Introduction.}
The scenario of ``unitary limit'' was originally formulated by Bertsch in 1999,
asking what will be the ground-state properties of a spin-1/2 fermion system
with an interaction of 
infinite scattering length \cite{Bertsch}.
With impressive advances of cold-atom experimental techniques,
such unitary Fermi systems became experimentally accessible at the atomic level,
and have attracted intensive attention.
This limit is also of great interest to nuclear systems,
because the $^1S_0$ scattering lengths of realistic nucleon-nucleon interactions
are all fairly large, such as $-$18.97 fm in the CD-Bonn potential~\cite{CDBonn},
in comparison with other length scales in the nuclear system.

For a Fermi gas at the unitary limit, the ground-state energy $E_0$
is expected to be proportional to the energy of the corresponding free gas $E_0^{\rm free}$,
i.e., $E_0=\xi E_0^{\rm free}$ with $\xi$ an universal constant shared by all unitary Fermi gases.
There have been huge efforts devoted to describing universal behaviors of unitary Fermi gases
in regard to the ground-state properties, as well as collective excitations, thermodynamic properties,
and non-equilibrium aspects, see e.g.~Refs.~\cite{Jr1999,Heiselberg2001,Carlson2003,Perali2004,Bulgac2005,Nishida2006,Haussmann2007,Schaefer2007,Siu2008,Dong2010,Schaefer2010,exp1,exp2,exp3,exp4,exp5,exp6,exp7,exp8,exp9}.
Our previous results of unitary neutron matter \cite{Siu2008,Dong2010}
by summing the low-momentum particle-particle hole-hole
ring diagrams to all orders,
with very different unitarity potentials, give $\xi$
values all closely equal to 0.44.
Very recently, extensive studies on few-nucleon systems
\cite{Konig2017,Kievsky2018} and neutron stars \cite{Kievsky2018}
from perspective of the unitary limit
have been carried out.

A unitary system of finite-number fermions
confined in a harmonic trap is
receiving increasing attention \cite{Werner2006,virial,
Chang2007,Blume2007,virial2,Yin2015}.
In previous works, see e.g.~\cite{Chang2007,Blume2007,Yin2015}, unitary systems
confined in a harmonic trap,
whose size $a_{\rm h.o.}$$=$$\sqrt{\hbar/(m\omega)}$
is much larger than the effective range $r_{\rm e}$ of the inter-fermion interaction, were studied.
The trapped unitary systems with $r_{\rm e}$
comparable to or larger than $a_{\rm h.o.}$, however,
remain largely unknown. In particular,
of great interest and importance is the question
whether or not such trapped systems at the unitary limit
are universally related to corresponding non-interacting ones.
In this work we find the ratio $E_0/E_0^{\rm free}$ for a
trapped unitary system with $r_{\rm e}$ comparable to or larger than $a_{\rm h.o.}$,
is remarkably invariant with variations of the harmonic trap parameter,
the number of fermions, the decimation momentum of the
low-momentum interaction
$V_{{\rm low}-k}$ \cite{vlowk1,vlowk2,vlowk3,vlowk4,vlowk5,vlowk6,vlowk7},
and the type and effective range of the inter-fermion potential,
suggesting a universal nature of these systems.
Our results support a special linear scaling relation of $E_0$,
which is shown to be analytically consistent with the unitary-limit virial theorems of Refs.~\cite{virial,virial2}.
We also study regularities of trapped unitary neutrons
from perspective of Landau's Fermi liquid theory \cite{MBT}.

{\bf Formalism.}
We consider a system of neutrons trapped
in a harmonic trap, with Hamiltonian $H=T+U_{\rm osc}+V$, where
$T$ represents the kinetic energy, $U_{\rm osc}$ the oscillator potential
$\sum_i \frac{1}{2}m\omega ^2 r_i^2$, and $V$ a unitary neutron-neutron
interaction which will be described later.
Let us consider a closed-shell system first.
The ground-state (g.s.)
energy $E_0$ of this system
will be calculated
using a linked-diagram expansion \cite{Goldstone,Kuo1971,Kuo1990} of the form
\begin{eqnarray}
E_0  &=& E_0^{\rm free} + \Delta E_0~, \nonumber \\
 \Delta E_0 &=&
     \frac{ \langle\Phi_0|V U(0,t')|\Phi_0\rangle}
         { \langle\Phi_0| U(0,t')|\Phi_0\rangle}|_{t'\rightarrow -\infty}~,
     \\
E_0^{\rm free}&=& \langle\Phi_0|U_{\rm osc}|\Phi_0\rangle
+\langle\Phi_0|T|\Phi_0\rangle ~\equiv~ U_0 + T_0~,  \nonumber
\end{eqnarray}
where $ U(t,t')$ is the time evolution operator. $E_0^{\rm free}$ is
the non-interacting g.s.~energy, and $\Delta E_0$ is given by the sum of
all the linked diagrams generated by the interaction $V$.
$\Phi _0$ is a closed-core wave function, such as a shell-model
state with the shells $(0s,0p,0d1s,0f1p)$
all filled when considering a system of 40 neutrons.
$U_0=T_0= \sum_{a\leq k_F} \frac{1}{2} \epsilon^{(0)}_a$,
where $\epsilon_a ^{(0)}$ is the oscillator single-particle (s.p.) energy of the orbit $a$,
and $k_F$ is the top filled s.p.~orbit.

We define a trap ratio $R_t$ for neutrons in a harmonic trap as
\begin{equation}
R_t = \frac{E_0}{E_0^{\rm free}}=\frac{U_0+T_0+\Delta E_0}{U_0+T_0}~. \label{Rt}
\end{equation}
We also define an intrinsic ratio $R_i$ as
\begin{equation}
R_i =\frac{T_0-K_{\rm cm}+\Delta E_0}{T_0-K_{\rm cm}}~. \label{Ri}
\end{equation}
Here $K_{\rm cm}$ is the kinetic energy of the center-of-mass (c.m.) motion,
and we have $K_{\rm cm}= \frac{3}{4}\hbar\omega$ since $\Phi_0$
is a shell-model closed-core state.
The ratio $R_i$ is based on the intrinsic kinetic energy of the neutrons
excluding that from their c.m. motion, and the interaction energy  generated
by the inter-neutron interaction $V$.
We shall study both ratios at the unitary limit, namely
as the scattering length of $V$ approaches infinity.

We employ two types of unitary interactions  in this work;
one is the high-precision meson-exchange CD-Bonn potential
\cite{CDBonn} and the other is a hard-core square well (HCSW)
potential \cite{Dong2010}, both tuned to have their
scattering lengths approaching infinity.
The tuning of the CD-Bonn potential is carried out
by slightly adjusting its $\sigma$ meson mass by $2.4\%$,
leading to
a CD-Bonn potential with its $^1S_0$
scattering length $a_s$=$-$12070 fm \cite{Siu2008},
which is obviously enormous compared to any other length
scales in the nuclear system.

The HCSW
interactions \cite{Dong2010}  are of the form
\begin{equation}
V(r)=V_c~(r\le r_c),~~V_b~(r_c < r \le r_b),~~0~(r > r_b),
\end{equation}
where $V_c,r_c,V_b$ and $r_b$ are parameters.
Since the scattering lengths of them can be
 given analytically \cite{Dong2010},
exact unitary HCSW interactions are readily obtained.
The five potentials employed in this work
(denoted as HCSW-1, 2, 3, 4, 5) all have their $V_c$=3000 MeV,
and their other parameters  $r_c$=(0.30, 0.15, 0.30, 0.50, 0.30) fm,
$V_b$=($-$20, $-$20, $-$30, $-$50, $-$50) MeV  and $r_b$=(2.41, 2.31, 2.03, 1.81, 1.62) fm.
We shall check whether or not the ratios are invariant
with respect to the type of the unitarity potential and with the variation of the effective range.

Next we obtain
a low-momentum interaction $V_{{\rm low}-k}$ from the above unitarity potential.
A renormalization procedure
\cite{Siu2008,vlowk1,vlowk2,vlowk3,vlowk4,vlowk5,vlowk6,vlowk7}
where the high-momentum
components of such a potential are integrated out is enacted.
Note that the resulting $V_{{\rm low}-k}$ preserves both the scattering length and the effective range.
For most cases, we shall use the decimation momentum
$\Lambda$=2.0 fm$^{-1}$
as used in a number of nuclear structure studies \cite{vlowk3}.
We shall also check whether the ratios are invariant with
the variation of $\Lambda$.

We calculate
the g.s.~energy shift $\Delta E_0$ using a ring-diagram method
(see Refs.~\cite{Siu2008,pphh1,pphh2} for details),
where the particle-particle hole-hole
($pphh$) ring diagrams are summed to all orders.
The low-order $pphh$ ring diagrams 
can be readily calculated.
The 1st- and 2nd-order diagrams are respectively
\begin{eqnarray}
D^{(1)} &=&  \frac{1}{2}\sum\limits_{ab} \langle ab|V_{{\rm low}-k}|ab
\rangle n_a n_b, \label{D1} \\
D^{(2)} &=&
 \sum\limits_{abcd}
\frac{\langle ab|V_{{\rm low}-k}|cd \rangle^2 n_a n_b (1-n_c)(1-n_d)}
{4\times [\bar\epsilon_a + \bar \epsilon_b -\bar \epsilon_c -\bar \epsilon_d]}.
  \label{D2}
\end{eqnarray}
Here $n_a$ is the occupation number of the orbit $a$,
given by $n_a$=1 if $a$$ \leq $$k_F$ and $=$0 if otherwise,
and $\langle ab | V_{{\rm low}-k} |cd \rangle$ denotes the
anti-symmetrized and normalized shell-model matrix element.
$\bar\epsilon$ is a shell-model Hartree-Fock (smHF) s.p.~energy.
We dress the single-particle and single-hole lines
with self-energies, or HF insertions,
and include such insertions to all orders.
Sample diagrams illustrating the general structure
of the $pphh$ ring diagrams and their self-energy
insertions have been given in e.g. Refs.~\cite{Siu2008,pphh1}.
The net effect of such all-order insertions
is the replacement of the oscillator
s.p.~energy $\epsilon^{(0)}$ of the undressed ring diagrams
by the smHF s.p.~energy 
\begin{equation}
\bar \epsilon_a= \epsilon ^{(0)}_a + \sum\limits_{b} \langle ab|V_{{\rm low}-k}|ab \rangle n_b~.
\end{equation}
Unless specified otherwise, we shall use $\bar \epsilon$
in all the calculations reported in this work.

An initial step to calculate the energy shift given by
the all-order sum of the $pphh$ ring diagrams (denoted as $\Delta E_0^{\rm pp}$),
is to solve the following RPA equation \cite{Siu2008,pphh1}.
\begin{eqnarray}
&&\sum\limits_{cd\in P} [ (\bar \epsilon_a+\bar \epsilon _b ) \delta_{ab,cd}
+ \lambda (1-n_a-n_b)  \nonumber\\
&& ~ \langle ab|V_{{\rm low}-k}|cd \rangle ] \times Y_k(cd,\lambda) =
\omega_k Y_k(ab,\lambda)~.  \label{RPAeq}
\end{eqnarray}
Here $\lambda$ is a strength parameter, to be varied from 0 to 1.
The indices $(a,b)$ can be either both particles $(p,p')$ or
both holes $(h,h')$ and so are $(c,d)$. The subscript $P$
refers to the ranges for particles and holes used in the
calculation. As an  example, for  the 40-neutron case
we have $(h,h')$ in the core composed of the
$(0s,0p,0d1s,0f1p)$ shells
and $(p,p')$ in the two major shells above the core.
The above equation has two sets of solutions
$(\omega_n^+,Y_n)$ and $(\omega_m^-,Y_m)$,
with $Y_n$ and $Y_m$ dominated respectively by their $(p,p')$ and $(h,h')$ components.
With the wave functions of the latter set, the energy shift is then given by \cite{pphh1}
\begin{eqnarray}
&&\Delta E_0^{\rm pp} = \int_0^1 d\lambda
 \sum\limits_m \sum\limits_{ab,cd\in P} Y_m(ab,\lambda) \nonumber \\
&& \times Y^*_m(cd,\lambda) \langle ab|V_{{\rm low}-k}|cd \rangle~. \label{eigen-vec}
\end{eqnarray}
The energy shift can also be calculated using a different
and considerably simpler method \cite{pphh2}, i.e.,
\begin{eqnarray}
\Delta E_0^{\rm pp}= -  \sum_m \omega_m^-
 + \sum_{ab} \left( \bar \epsilon _a n_a + \bar \epsilon _b n_b \right)~, \label{eigen-val}
\end{eqnarray}
where $\omega_m^-$ is obtained by solving Eq.~(\ref{RPAeq}) with $\lambda$=1.
We shall carry out our calculations
using both methods to cross check our numerical results.

\begin{table*}
\centering
\caption{\label{table1} The trap ratio $R_t$ for two closed-shell systems with $A$=40 and 70,
as well as the intrinsic ratio $R_i$ for $A$=70. The ratios given by
summing up the $pphh$ ring diagrams
to all orders and the 1st- and 2nd-order approximations are denoted respectively
as $R^{({\rm all})}$, $R^{(1)}$ and $R^{(2)}$.
The decimation momentum $\Lambda$ of the $V_{{\rm low}-k}$
is in units of fm$^{-1}$,  the effective range $r_{\rm e}$ of the inter-neutron potential
and the harmonic trap size $a_{\rm h.o.}$=$\sqrt{\hbar/(m\omega)}$ in fm, and $\hbar \omega$ in MeV.}
\begin{spacing}{1.3}
\begin{tabular}{ccccccccccccccccccccccccccccccccccccccccccccccccccccccccccccccccccccccccccccccccccccccccccccccccccccccccccccccccccc}
\hline \hline

           & \multirow{2}{0.25cm}{$\Lambda$} & \multirow{2}{0.3cm}{$r_{\rm e}$}  & \multirow{2}{1.3cm}{$a_{\rm h.o.}$/$\hbar \omega$}
                                    &     \multicolumn{3}{c}{40 neutrons}       &&    \multicolumn{3}{c}{70 neutrons}        \\ \cline{5-7} \cline{9-11}
           &&&                      &$R_t^{(1)}$           &$R_t^{(2)}$           &$R_t^{({\rm all})}$
                                   &&$R_t^{(1)}/R_i^{(1)}$&$R_t^{(2)}/R_i^{(2)}$&$R_t^{({\rm all})}/R_i^{({\rm all})}$     \\ \hline

CD-Bonn   ~&~ 2.0 ~&~ 2.54 ~&~2.35/7.5 ~&~ 0.763 ~&~ 0.758 ~&~ 0.756 ~&&~ 0.754/0.505 ~&~ 0.752/0.502 ~&~ 0.752/0.501     \\

CD-Bonn   ~&~ 2.0 ~&~ 2.54 ~&~2.21/8.5 ~&~ 0.760 ~&~ 0.755 ~&~ 0.754 ~&&~ 0.752/0.502 ~&~ 0.751/0.499 ~&~ 0.750/0.498     \\

CD-Bonn   ~&~ 2.0 ~&~ 2.54 ~&~2.09/9.5 ~&~ 0.759 ~&~ 0.754 ~&~ 0.753 ~&&~ 0.752/0.502 ~&~ 0.751/0.499 ~&~ 0.750/0.498     \\

CD-Bonn   ~&~ 2.0 ~&~ 2.54 ~&~1.99/10.5 ~&~ 0.758 ~&~ 0.754 ~&~ 0.754 ~&&~ 0.753/0.503 ~&~ 0.752/0.501 ~&~ 0.751/0.500     \\

CD-Bonn   ~&~ 2.0 ~&~ 2.54 ~&~1.86/12.0 ~&~ 0.759 ~&~ 0.756 ~&~ 0.755 ~&&~ 0.755/0.508 ~&~ 0.754/0.506 ~&~ 0.754/0.506     \\

CD-Bonn   ~&~ 2.0 ~&~ 2.54 ~&~1.72/14.0 ~&~ 0.762 ~&~ 0.760 ~&~ 0.759 ~&&~ 0.761/0.519 ~&~ 0.760/0.518 ~&~ 0.760/0.517     \\

CD-Bonn   ~&~ 2.3 ~&~ 2.54 ~&~1.99/10.5 ~&~ 0.758 ~&~ 0.754 ~&~ 0.754 ~&&~ 0.753/0.503 ~&~ 0.752/0.501 ~&~ 0.751/0.500     \\

CD-Bonn   ~&~ 1.8 ~&~ 2.54 ~&~1.99/10.5 ~&~ 0.758 ~&~ 0.754 ~&~ 0.753 ~&&~ 0.752/0.502 ~&~ 0.751/0.500 ~&~ 0.751/0.499     \\

HCSW-1    ~&~ 2.0 ~&~ 2.63 ~&~1.99/10.5 ~&~ 0.761 ~&~ 0.759 ~&~ 0.759 ~&&~ 0.762/0.521 ~&~ 0.761/0.520 ~&~ 0.761/0.519     \\ 

HCSW-2    ~&~ 2.0 ~&~ 2.36 ~&~1.99/10.5 ~&~ 0.754 ~&~ 0.751 ~&~ 0.750 ~&&~ 0.750/0.498 ~&~ 0.749/0.496 ~&~ 0.749/0.495     \\ 

HCSW-3    ~&~ 2.0 ~&~ 2.22 ~&~1.99/10.5 ~&~ 0.753 ~&~ 0.750 ~&~ 0.749 ~&&~ 0.749/0.495 ~&~ 0.748/0.493 ~&~ 0.747/0.492     \\ 

HCSW-4    ~&~ 2.0 ~&~ 2.20 ~&~1.99/10.5 ~&~ 0.753 ~&~ 0.749 ~&~ 0.748 ~&&~ 0.749/0.496 ~&~ 0.748/0.493 ~&~ 0.748/0.493     \\ 

HCSW-5    ~&~ 2.0 ~&~ 1.80 ~&~1.99/10.5 ~&~ 0.760 ~&~ 0.754 ~&~ 0.753 ~&&~ 0.750/0.498 ~&~ 0.748/0.494 ~&~ 0.748/0.493     \\ 

\hline
\end{tabular}
\end{spacing}
\end{table*}

\begin{figure}
\centering
\includegraphics[width = 0.40\textwidth]{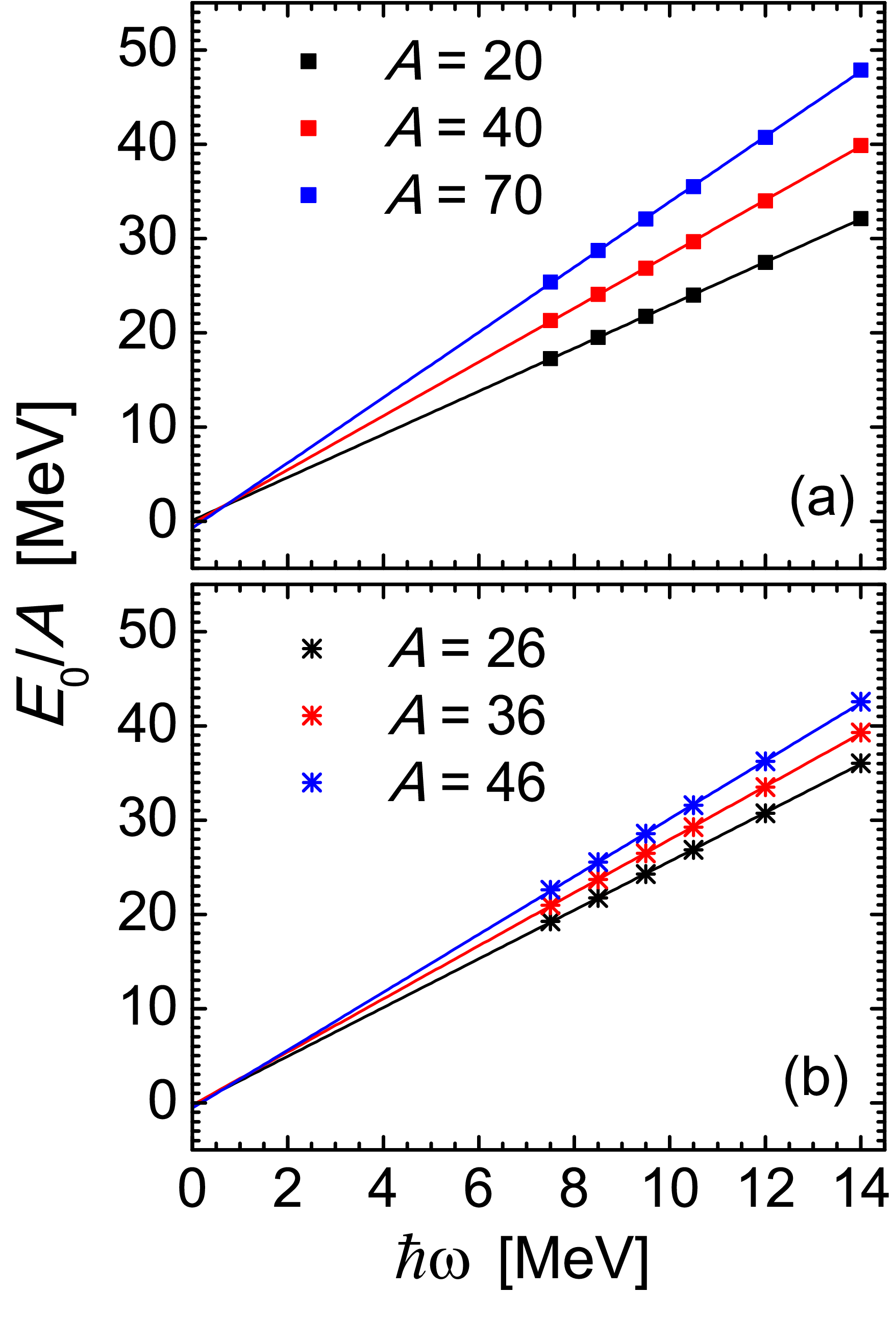}
\caption{\label{fig2} Energy per neutron (denoted as $E_0/A$)
versus the $\hbar\omega$ value, for (a) closed-shell systems and (b) open-shell systems.
See text for further explanations.}
\end{figure}

\begin{figure}
\centering
\includegraphics[width = 0.42\textwidth]{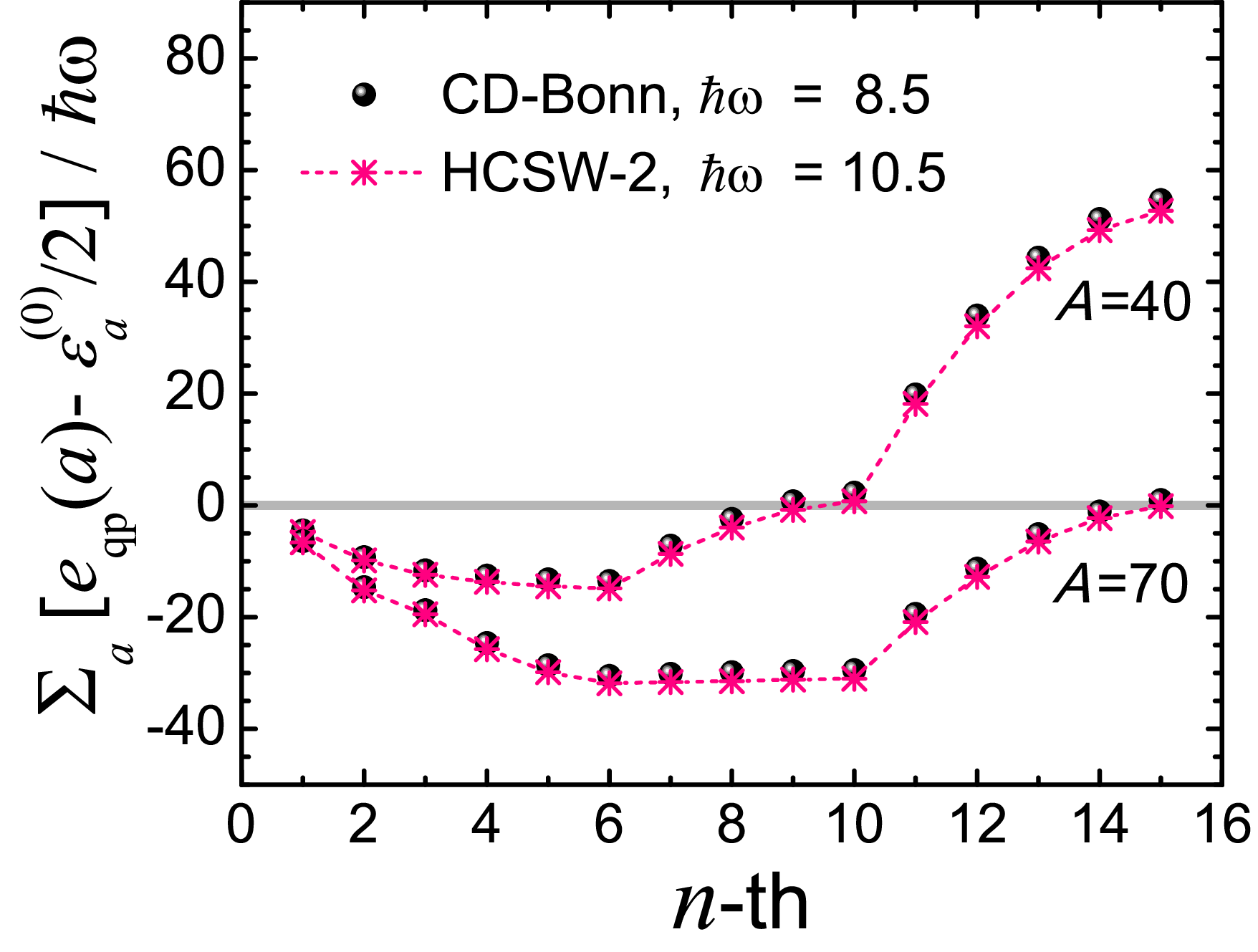}
\caption{\label{fig6} Summation of $(e_{\rm qp}(a)-\frac{1}{2}\epsilon _a ^{(0)})$
(equal to intrinsic quasi-particle energies $e^{\rm int}_{\rm qp}(a)$) up to the $n$-th orbit,
for two closed-shell systems with $A$=40 and 70, respectively.
The values over $\hbar\omega$ are presented here.
See text for further explanations.
}
\end{figure}

\begin{figure}
\centering
\includegraphics[width = 0.40\textwidth]{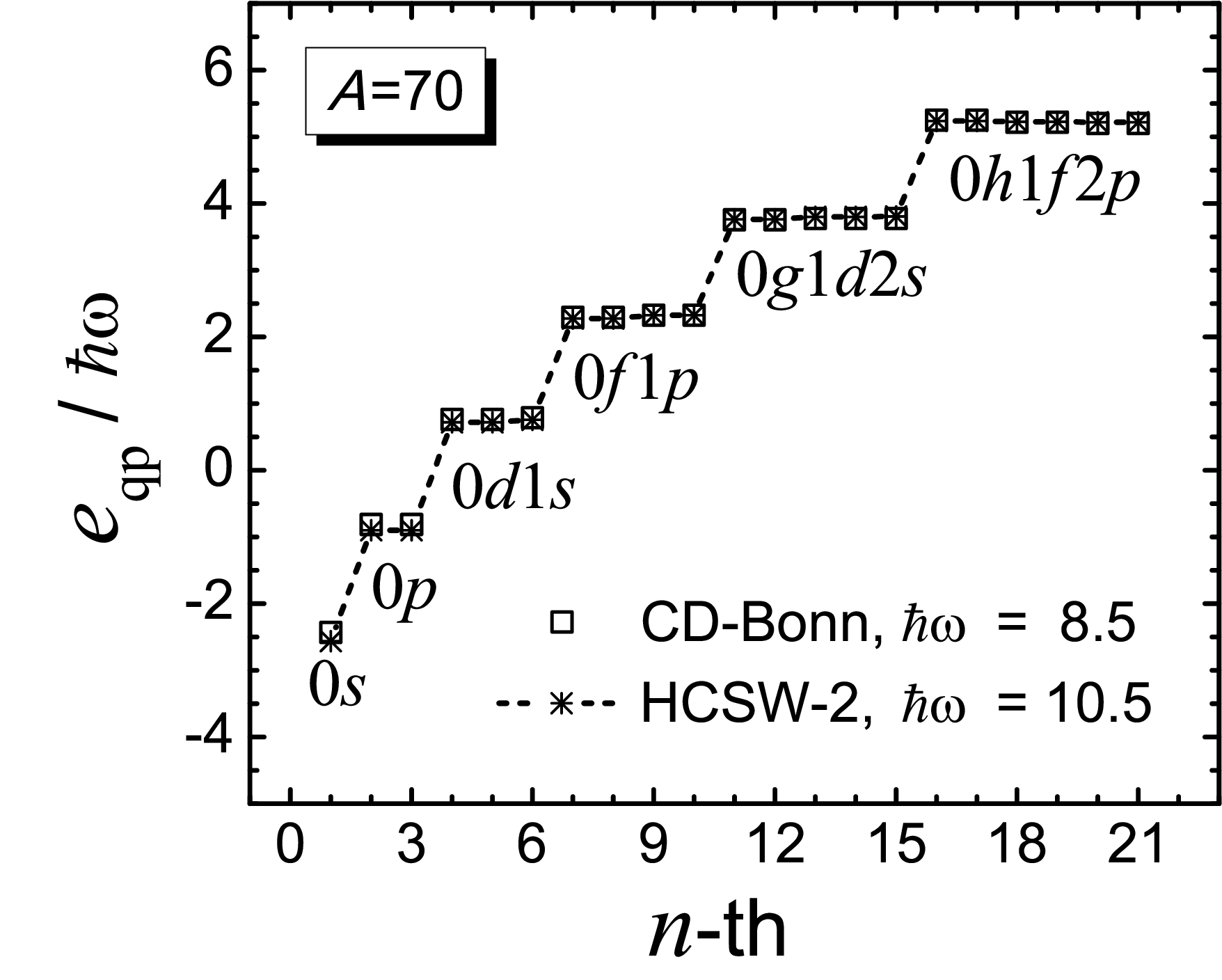}
\caption{\label{fig7} The quasi-particle energy $e_{\rm qp}$ of the $n$-th orbit,
for the closed-shell system with $A$=70.
The values over $\hbar\omega$ are presented here. See text for further explanations.
}
\end{figure}

{\bf Results and discussions.}
In this section we present and discuss the calculated g.s.~properties
for finite neutron systems confined in a harmonic trap
interacting with unitary interactions.
In Table \ref{table1} we present the trap ratio $R_t$ of Eq.~(\ref{Rt}) for two closed-shell systems with $A$=40 and 70,
as well as the intrinsic ratio $R_i$ of Eq.~(\ref{Ri}) for $A$=70.
In fact the ratios $R_t$ and $R_i$ are simply related by $2R_t= (1+R_i)$, if we neglect
$K_{\rm cm}$ in Eq.~(\ref{Ri}) which is obviously small compared with $T_0$ for our cases.
The ratios with the superscript ``all'' are given
by the g.s.~energy shifts calculated by
summing the $pphh$ ring diagrams
to all orders, using both the method of Eq.~(\ref{eigen-vec})
and that of Eq.~(\ref{eigen-val}).
The results given by the two methods are identical (to fourth decimal),
providing a check of our numerical calculations.
The ratios with the superscripts ``1'' and ``2'' are given
by the 1st- and 2nd-order approximations, whose energy shifts
are respectively $D^{(1)}$ and $(D^{(1)}+D^{(2)})$ of Eqs.~(\ref{D1}, \ref{D2}).

In Table \ref{table1} one sees that, 
starting from the 1st-order approximation,
the ratios decrease slightly by including the 2nd-order diagram.
The ratios given by the 2nd-order approximation are further decreased,
only very slightly, by summing the diagrams to all orders.
These very small differences indicate that  our  linked-diagram expansion
provides  a rapidly converging framework for calculating
the g.s. energies of trapped closed-shell unitary systems.
As also shown in Table \ref{table1}, of great interest is that
the $R_t^{\rm (all)}$ (and $R_i^{\rm (all)}$) values
are remarkably invariant in regard to variations of the trap parameter $\hbar\omega$,
the decimation momentum $\Lambda$ of $V_{{\rm low}-k}$,
and the type (either the meson-exchange CD-Bonn potential or the HCSW ones)
and effective range $r_{\rm e}$ of the unitarity potential.

Below we study a connection between
our results and the unitary-limit virial theorems \cite{virial,virial2}.
Let us 
begin with the theorem
of Werner and Castin \cite{virial}.  According to this theorem,
the g.s.~energies of unitary fermion systems in a harmonic trap
satisfy the relation
\begin{equation}
\langle \Psi_0|[T +U_{\rm osc} +V]|\Psi _0   \rangle=E_0
=2 \langle \Psi_0|U_{\rm osc}|\Psi _0   \rangle \,,  \nonumber
\end{equation}
where
$E_0$ and $\Psi_0$ are respectively the g.s.~energy
and wave function of the system.

From the above equation we readily have
\begin{equation}
\omega \frac{d}{d\omega}E_0=E_0~,   \label{v1}
\end{equation}
which gives
\begin{equation}
 E_0=\alpha \hbar \omega  \label{linear}
\end{equation}
with $\alpha$ a constant independent of $\hbar \omega$.
Eq.~(\ref{linear}) is a special linear scaling relation
for the g.s.~energy 
of a unitary fermion system in a harmonic trap.
Let us  check if our calculated
g.s. energies are consistent with this scaling relation.


In Fig.~\ref{fig2}(a) we present our ring-diagram
g.s.~energies using the fine-tuned CD-Bonn potential
and using $\hbar\omega$=7.5, 8.5, 9.5, 10.5, 12.0, 14.0 MeV,
respectively.
Three closed-shell systems with $A$=20, 40, and 70 are considered.
Plotted is the energy per neutron (dots),
versus the $\hbar\omega$ value,
as well as the corresponding linear-fitting result (line).
One sees all lines fit the data nearly perfectly,
and moreover   they all converge to
the origin $(E,\hbar\omega)$=$(0,0)$
also near perfectly. Clearly our results are in highly
satisfactory agreement with the linear scaling relation
of Eq.~(\ref{linear}).
Returning to Table \ref{table1}, the ratios $R_t$
presented there are practically
invariant with the variation of $\hbar \omega$.
This invariance is actually
a consequence of the above scaling relation
(recalling $R_t=E_0/E_0^{\rm free}$ and $E_0^{\rm free}$ is proportional to $\hbar\omega$).

We also generalize our study to open-shell systems
with valence neutrons occupying the major shell just
above the closed core.
Based on the low-momentum unitary interactions the same as those
used for closed-shell systems,
we derive the one-body and two-body effective interactions
for valence neutrons
using a $\hat{Q}$-box folded-diagram method \cite{vlowk3,Kuo1990}.
Here the $\hat{Q}$-box vertex function is composed of
1st- and 2nd-order irreducible valence-linked diagrams,
including the core-polarization ones.
We then calculate and diagonalize the Hamiltonian matrix
in the many-body model space of valence particles,
using a standard shell-model code \cite{nushellx}.
In Fig.~\ref{fig2}(b)
we exemplify the results of open-shell systems
using those with $A$=26, 36, and 46.
One sees the energy per
neutron is also remarkably proportional to $\hbar\omega$,
again in very good agreement with the scaling relation of Eq.~(\ref{linear}).
According to our calculation, the scaling relation also
holds perfectly for odd-$A$ open-shell systems.

The g.s.~energies $E_0$
of Fig.~\ref{fig2} are calculated with finite effective range
$r_{\rm e} \simeq a_{\rm h.o.}$ (see Table \ref{table1}),
and they are in good agreement
with the linear scaling relation of Eq.~(\ref{linear}) which
is, however, based on the virial theorem \cite{virial} for
$r_{\rm e}$$\rightarrow$0. There is a difference in $r_{\rm e}$.
In Ref.~\cite{virial2}
a general virial theorem including the effect of
finite $r_{\rm e}$ was obtained.
From this theorem,
we have
\begin{eqnarray}
\omega \frac{d}{d\omega}E_0 = E_0 + \frac{1}{2}(E_0-\bar{E}_0)~,   \label{v2}
\end{eqnarray}
where $E_0$ and $\bar{E}_0$ are the g.s.~energies of systems
with $r_{\rm e}$$\neq$0 and =0, respectively.
As illustrated in Table \ref{table1},
the ratios $R_t$ is invariant with the choice of the unitary
interaction, and consequently $E_0$ is invariant with $r_{\rm e}$
(recalling $R_t=E_0/E_0^{\rm free}$ and $E_0^{\rm free}$ is independent of $r_{\rm e}$).
Thus we have in general $E_0= \bar{E} _0$, and the above differential equation
becomes $\omega (d/d\omega) E_0=E_0$
which is identical to Eq.~(\ref{v1}).
This leads to an interesting conclusion, 
namely the scaling relation
of Eq.~(\ref{linear}) is applicable to not only trapped unitary systems with
$r_{\rm e}$$\rightarrow$0 but also those with $r_{\rm e}$$>$0, 
which is supported by
our results in Fig.~\ref{fig2}.

At last we study the unitary ratio guided by Landau's Fermi liquid (FL) theory \cite{MBT}.
In this way, a simple relation between the ratio
and the Fermi-liquid quasi-particles can be obtained.
For a system of neutrons in a harmonic trap, the quasi-particle (q.p.)
energy $e_{\rm qp}$ and the g.s.~energy $E_0$ are related as
\begin{eqnarray}
e_{\rm qp}(a) &=& \frac{\delta E_0}{\delta n_a}~, \nonumber \\
E_0 &=& \frac{1}{2}[\sum _a \epsilon_a ^{(0)}n_a + \sum _a e_{\rm qp}(a) n_a]~.
\end{eqnarray}
The trap ratio of Eq.~(\ref{Rt}) is then given by $\sum _a [e_{\rm qp}(a)-\frac{1}{2}\epsilon _a ^{(0)} ]n_a = 2(R_t-\frac{3}{4})E_0^{\rm free}$.
To have $R_t$=0.75, the FL quasi-particles should satisfy the sum rule
\begin{equation}
\sum\limits_a [e_{\rm qp}(a)-\frac{1}{2}\epsilon _a^{(0)}] n_a = 0~. \label{sumrule}
\end{equation}
We also define an intrinsic q.p.~energy $e^{\rm int}_{\rm qp}(a)=e_{\rm qp}(a)-\frac{1}{2}\epsilon _a^{(0)}$,
and the intrinsic ratio of Eq.~(\ref{Ri}) is then given by $\sum _a e^{\rm int}_{\rm qp}(a) n_a = 2(R_i-\frac{1}{2})T_0$ if $K_{\rm cm}$ is neglected.
One easily sees the sum rule also gives $R_i$=0.5.

As shown in Table \ref{table1}, the g.s.~energy given by the low-order approximation is closely equal to that including high-order diagrams.
We have thus approximated $e_{\rm qp}$ by $e_{\rm qp}^{(2)}(a) = \epsilon _a^{(0)} + \delta[D^{(1)}+D^{(2)}] / \delta n_a$,
and obtain $e_{\rm qp}^{(2)}$ by solving this equation in a self-consistent manner.
In Fig.~\ref{fig6} we exemplify the above sum rule, using the closed-shell systems with $A$=40 and 70.
The results are based on the unitary CD-Bonn and HCSW-2 interactions,
combined with the trap parameter $\hbar\omega$=8.5 and 10.5 MeV, respectively.
As shown in Fig.~\ref{fig6}, the sum rule of
Eq.~(\ref{sumrule}) is indeed well satisfied by our results.
Moreover, the scaled q.p.~energies $e_{\rm qp}$/$\hbar\omega$
are remarkably invariant with the choice of the unitary interaction
and the choice of $\hbar \omega$,
as shown in Fig.~\ref{fig6} and later in Fig.~\ref{fig7}.
In Fig.~\ref{fig7} we present the q.p.~energies of the $A$=70 system,
where one sees the $e_{\rm qp}$ values for orbits
within one harmonic oscillator
shell are nearly degenerate.
According to our calculation, the q.p.~energies of $A$=20 and 40
also have this major-shell degeneracy.
In ordinary nuclear systems the s.p.~energy levels
of each major shell are generally non-degenerate.
It will be of much interest to check if this drastic change does take place experimentally.

{\bf Summary.}
We have studied systems of finite-number neutrons in a harmonic
trap at the unitary limit.
Two very different types of neutron-neutron interactions have
been applied, namely, the
meson-theoretic CD-Bonn potential and the hard-core square-well ones, all
tuned to possess infinite scattering lengths,
and with effective ranges comparable to or larger than the trap size.
The potentials were renormalized
to equivalent, scattering-length preserving
low-momentum potentials, $V_{{\rm low}-k}$,
with which the particle-particle hole-hole
ring diagrams are summed to all orders
to yield the ground-state energy $E_0$ of the finite neutron system.
Two different methods were employed
for the above ring-diagram calculations,
giving practically identical results.
We find the ratio
$R_t \equiv E_0/E_0^{\rm free}$
to be remarkably invariant in regard to variations of the harmonic
trap parameter,
the number of neutrons,
the decimation momentum of $V_{{\rm low}-k}$,
and the type and effective range of the unitarity potential.
Our results support a special linear scaling relation of $E_0$,
which is shown to be analytically consistent with the unitary-limit virial theorems.
Our results further suggests the scaling relation is applicable to
not only trapped unitary systems with inter-fermion interactions of zero-range
but also those with interactions of finite effective ranges.
The $R_t$'s all flow to a specific value of 0.75, 
suggesting a sum rule of Landau's quasi-particles for trapped unitary neutrons.
The quasi-particle energies also exhibit a major-shell degeneracy behavior.

{\bf Aknowledgments:}
We thank Ismail Zahed and Jeremy W. Holt for many helpful discussions.
The works by T.T.S.K. and R.M. were supported in part by the U.S. Department of Energy
under Award Numbers DE-FG02-88ER40388 and DE-FG02-03ER41270, respectively.
Y.Y.C. and Y.M.Z. thank the National Natural Science Foundation of China (Grant Nos.~11875134, 11505113 and 11675101),
and the Program of Shanghai Academic/Technology Research Leader (Grant No.~16XD1401600) for financial supports.

\end{document}